# Routing of Electric Vehicles in a Stochastic Network with Non-recurrent Incidents

Mohammad Arani[*]
*Dept. of Systems Engineering*
*University of Arkansas at Little Rock*
Little Rock, AR, USA
mxarani@ualr.edu

Mohammad Mehdi Rezvani
*Dept. of Electrical Engineering*
*Louisiana State University*
Baton Rouge, LA, USA
mrezva2@lsu.edu

Hamzeh Davarikia
*Dept. of Electrical Engineering*
*McNeese State University*
Lake Charles, LA, USA
hdavarikia@mcneese.edu

Yupo Chan
*Dept. of Systems Engineering*
*University of Arkansas at Little Rock*
Little Rock, AR, USA
yxchan@ualr.edu

---

[*] Corresponding author.

*Abstract*—An approach for mapping an electric vehicle (EV) driver's travel time constraints and risk-taking behavior to real-time routing in a probabilistic, time-dependent (or stochastic) network is proposed in this paper. The proposed approach is based on a heuristic algorithm that finds the shortest path according to the driver's preferences. Accounting for en-route delays and alternate routes, the EV routing problem in stochastic networks is shown to exhibit other than the First-In-First-Out (FIFO) property; i.e., the traveling time for those who depart earlier may not sooner of those who depart later or wait en-route in the charging stations. The proposed approach provides EV drivers the option to manage their trip and reach the destination on time, while by taking advantage of the non-FIFO characteristics of the traffic network, charge their cars en-route. The proposed routing algorithm is tested on a given stochastic transportation network. The best routes based on the driver's preferences are identified while accounting for the best-planned delays at the charging stations or en-route.

1. INTRODUCTION

In today's highway network, mobility is on the increase and infrastructure supply has been overwhelmed by travel demand. This necessitates optimizing the use of existing infrastructures, including the use of advanced traveler information system (ATIS). A piece of information of paramount importance to travelers is the en-route travel time and routing options, given that traffic conditions and incident risks change frequently. The routing problem (RP) for conventional vehicles has a mature history. Hu & Chan [1] propose an algorithm which allows waiting en-route and avoiding incident risks. By waiting en-route, it is inferred that a driver stays in a queue patiently, rather than balking. Here, the difference between "deliberate/voluntary" waiting (when there is a chance to balk) and "forced" waiting (for an unexpected delay during traffic congestion) is distinguished. "Waiting en-route" is defined as travelers following a congested route (thus waiting in the queue) instead of taking a detour along an alternative (uncongested, free-flowing) route. With electric vehicles (EV), "waiting en-route" entails waiting at charging stations. While waiting at the origin is typically most convenient, the driver of an EV can take advantage of waiting en-route by visiting a charging station to charge the EV battery.

The optimal routing for EVs along with locating charging stations is proposed in [2]. In this approach, the vehicle loading capacity and the customer time windows are considered, and a heuristic approach is employed to solve the problem. Bozorgi et al. [3] propose a new algorithm to solve the EV RP, which results in an extended driving range and battery durability. This approach, which benefits from historical driving datasets, employ data mining techniques to evaluate the time and energy required in the routes. Barco et al. [4] study the difference between RP for conventional and EVs. It is discussed that EVs can be viewed as a special case for the RP, since the autonomy, battery degradation, and the charging process of the EVs are involved.

As far as the network properties in RP is concerned, first in first out (FIFO) and non-FIFO properties need special attention.

In a network with FIFO property, it is generally assumed that a commuter who departs the origin earlier will arrive the destination earlier [5]. However, with non-FIFO property, this assumption does not necessarily hold. As discussed in [6], for dynamic and stochastic shortest path problem (DSSPP) in ATIS, the future traffic uncertainty must be taken into account by using forecast data when the travel time is to be estimated; a consideration that leads to the non-FIFO property.

Chan et al. [7] proposed an algorithm which implicitly detects bottlenecks in a stochastic network, the algorithm recommends the best routing and waiting time leading a driver reach to her destination earlier than the one who departs ahead of her. Although the result appears counter-intuitive, the Bellman's optimality condition is met in the real-world time-dependent network of Central Arkansas Highway network, noting that the observed network entails time-dependency and risk of taking a route. The Decreasing Order of Time (DOT) and Waited-Search DOT (WSDOT) algorithms utilized in this paper reveals a promising result in the routing problem. Furthermore, the proposed algorithm can be implemented for helping the power grid by incorporating the proposed methods [8, 9] in as its constraints.

2. CONTRIBUTIONS AND NOVELTY

In this paper, a stochastic network is characterized not only by time-dependent travel times but also uncertainties due to unexpected, paroxysmal delays, which could be due to incidents. Unexpected delays here relate to probabilities of traffic breakdown and capacity reductions due to incidents. These incidents are assumed to be disseminated real-time to motorists by the ATIS, and an incident is associated with or concomitant with congestion delays. In the proposed algorithm, such real-time information is supplemented by historical safety records and the fundamental diagram of traffic flow (FDTF), which provides the probability of an incident at each arc [4, 10]. These probabilities are then inserted into the model and accounted for in route guidance. The proposed heuristic algorithm searches for the optimal next-hop node and optimal wait-time recursively, which can be considered as a more effective extension of the Hu-Chan algorithm in [1]. The model proposed in [1] demonstrates that travelers who stay on the congested route can arrive earlier at their destination than those who take a detour to presumably bypass the congestion. This "non-FIFO" property can be viewed from both a mesoscopic and microscopic traffic-theory perspective. The argument in microscopic traffic-flow theory was proposed in [11] and been called the Smeed's paradox ever since. For an arc, Smeed claimed that should disturbances propagate downstream, a trailing driver in a platoon could arrive at his destination earlier by leaving later. This notion was dismissed in [12], where the erroneous assumption of downstream propagation was pointed out based on the fact that disturbances can only propagate upstream.

The focus of this paper is on a mesoscopic rather than a microscopic traffic phenomenon. Generalizing from a path consisting of a series of arcs in a network, it is argued that the non-FIFO property can be exhibited in any network in which queuing is implicit—where drivers can follow more than one route from the origin to the destination similar to a queuing network. Notice that FIFO is applied here to allow for selecting routes, as assumed in the dynamic traffic assignment (DTA) literature [13]. It is widely accepted that FIFO property does not apply to a route level for which user-equilibrium (where travel time for each origin-destination pair is minimized) does not hold.

In addition to providing an EV the fastest route, the driver's travel time choices are evaluated in the proposed routing algorithm, as well. The evaluations are carried out in terms of a risk metric, which is defined as the probability of an incident occurring at a particular time of the day. By taking this probability into consideration, the driver's risk attitude is explicitly taken into account in of his/her routing policy. The risk metric for a particular arc, which is derived from historical safety records, indicates the potential for secondary incidents, which might occur as a result of a reported incident. Such consideration (all calculated on a real-time basis) lead not only to identifying the fastest route but also the most risk-free one.

### 3. TRANSPORTATION NETWORK

The stochastic nature of the transportation network is evident in our daily commute. For example, the travel time from home to work for different days of the weeks and different times of the day could vary significantly due to the disruptions arising from multiple random sources, such as incidents, vehicle breakdown, bad weather, work zones, special events, and etc. ATIS technology addresses some of these problems by supplying travelers with real-time traffic information on the entire transportation network. Fig. 1 shows an instance of a network equipped with the ATIS.

The goal of ATIS is to provide travelers with information that facilitates decisions concerning route choice, departure time, trip delay or elimination, and mode of transportation. Fig. 1 shows a driver at node 2 that is given guidance by the variable message signs (VMS) to what route to take. Solving these types of problems that drivers face in daily commute is one of the key needs for an ATIS system. While travel time is of importance in EV RP, risk of a hazard in the route and energy consumption are other factors that must be taken into account. The next two paragraphs discuss risk avoidance in traffic routing.

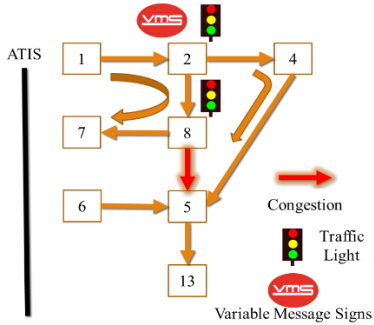

Fig. 1. A transportation network equipped with the ATIS

In a typical highway network, undesirable events do occur on some routes. Many are in the form of accidents, which might occur in poor road condition, near work-zones, and during special events, often resulting in a chaotic environment. Given that "risk-prone" routes can be obtained from historical data, risk-prone times of the day can also be determined for those routes. Usually, there is a higher frequency of incidents during peak traffic hours. In addition to recurring congestion, non-recurrent incidents are considered in this paper, as well. Thus, the risk is considered as a key factor in route choice, accounting for the chance of a driver to encounter undesirable conditions. In the forthcoming case studies, the risk is modeled as a time-dependent probability that is a function of the incident rate obtained from historical data.

### 4. ROUTING PROBLEM CONSIDERING THE ELECTRIC VEHICLES CHARACTERISTICS

The EV drivers require visiting the charging stations to charge the battery. With the current technology, charging takes longer time compared with fueling the conventional cars with internal combustion engines. Although this feature might be seen as a drawback, the proposed routing algorithm is designed to take advantage of this feature by using the en-route waiting time at the charging stations. To this end, the energy consumption and the state of the charge need to be modeled. Equation (1) shows that the total consumed energy, $w_{ij}$, for traversing arc $y_{ij}$, such that all arcs belong to a path from origin to destination, should be less than or equal to the remaining charge of the battery, $C$.

$$\sum_{(i,j)\in path} w_{ij} y_{ij} \leq C \qquad (1)$$

Considering the travel time, risk and energy consumption, the routing must be planned for a stochastic network where each arc has three attributes and the travel time is updated in a regular interval. To this end, we propose three routing algorithms, based on the FIFO and Non-FIFO decision making approaches that find the best route from the origin to the destination, considering the characteristics of the electric vehicles and the risk associated with the road trip. In the first step (initialization), all the possible routes from the source to the destination are found and the associated commuting time is calculated. The routes are then sorted based on the travel time in the descent order, and the driver preference is asked to accommodate the route selection for the risk-averse and risk-prone drivers. Every algorithm has been discussed in the following sub-sections.

*A. Algorithm 1- FIFO*

Algorithm 1 is associated with a driver whose mindset is FIFO, if he departs early, he will arrive early, and he is not willing to consider any stop node apart from recharging his battery. In this algorithm, no stop is planned during the traveling period.

| **Algorithm 1** Electric Vehicle Routing in the FIFO network |
|---|
| **(initialization):** |
| • t=0 (traffic is updated in 5-minute intervals) |
| • Find all possible routes from the origin to the destination. |
| • Calculate the travel time, energy consumption, and risk for each route. |
| • Sort the routes based on the travel time, energy consumption, and risk. |

- Acquire the driver's preference and choose one of the paths.

**(Main Body)**:

1) Start from origin
2) Find all emanating nodes and calculate the energy consumption to reach to those nodes by considering time intervals. At the same time calculate the shortest time to reach to each charging stations from those emanating nodes.
3) Sort out the nodes based on (1) travel time, and (2) the feasibility of reaching to the closest charging station from those.
    a. If there is no feasible charging station in the vicinity which means the EV stops at some point in the middle of the arc, there is no feasible solution for the problem. Go to the end.
4) Select the first from the sorted nodes and go to it. Update the battery level.
5) If there is a charging station:
    a. Just charge the battery as much as to reach to the next closest node and closest charging station calculated by sorting nodes set.
6) Go to number 2 until you reach your destination.

---

*B. Algorithm 2- Decreasing Order of Time (DOT)- FIFO*

The second algorithm is a modified algorithm adapted from Decreasing Order of Time (DOT) [7] to consider an Electric Vehicle's battery consumption. This algorithm utilizes the FIFO behavior of the network and does not consider any en-route.

The DOT algorithm which is first proposed by Chabini et. al [15] was used in implementing Dijkstra's shortest-path algorithm [16] for a time-dependent network. In DOT, the rationale behind the algorithm is to set a time when a driver needs to be at her destination. Modestly, use the concept of backward calculation given the set-time from destination to the origin in the time-dependent network and find the best routing. Therefore, the best commutation time is offered and the condition of arrival time is met, as well. Following the DOT algorithm, the proposed Algorithm 2 takes the path in the opposite direction of arcs, as it is described in the detail below.

**Algorithm 2** Decreasing Order of Time (DOT)--FIFO

**(initialization)**:

- t=0 (traffic is updated in 5-minute intervals)
- Find all possible routes from the origin to the destination.
- Calculate the travel time, energy consumption, and risk for each route.
- Sort the routes based on the travel time, energy consumption, and risk.
- Acquire the driver's preference and choose one of the paths.

**(Main Body)**:

1) Start from the destination node and set a level of charge for the battery at this node. (Minimum level will suffice)
2) Find the set of nodes terminating to the current node. Considering time dependency of the network. Calculate the time takes to commute those arcs and the battery level at those nodes. (higher battery level)
3) If the battery level is higher than max level, that arc is not feasible
4) If there is a charging station, reduce the battery level to the min level.
5) Sort out the nodes based on the time and feasibility of the arc-related to battery
6) Select the first from the sorted nodes and go to it. Update the battery level.
7) Go to number 2 until you reach your destination.

---

*C. Algorithm 3- Waited-Search DOT (WSDOT)- Non-FIFO*

The latest algorithm is the modified version of Waited-Search DOT (WSDOT) [7] algorithm that creates harmony among charging the battery and intended to wait at suggested nodes. In short, the WSDOT algorithm is like DOT algorithm and the difference is that the algorithm decides where and for how long wait at a node. Therefore there are two questions need to be addressed: (1) where to stop? (2) how long to wait at that node?

Question number one could be answered this way that we need to stop at a node if (a) we are allowed to stop, means, because we are proceeding by decreasing order of time, the time we reach to the origin should be $t \geq 0$, and the changes in time interval allow us to have a shorter travel time. Question number two could be answered by taking into consideration the same criterion. The waiting is as long as the time to change the interval and have a better trip time. The last point but not least is that to reach to the destination could be considered in two different scenarios as follow:

A. The first scenario is to compare WSDOT and DOT with the first algorithm (FIFO behavior), therefore the initialization for the latest time to reach is the best time obtained by the FIFO algorithm.

B. In the second scenario, the total trip time is important. Therefore we can consider a greater value for the initialization time, in order to just compare total trip time.

After introducing the routing algorithms, the next section demonstrates a case study considering the proposed approaches.

5. CASE STUDY

In order to validate the proposed routing algorithms, numerical analysis is carried out on travel time data gathered during peak and off-peak traffic hours on highway network of central Arkansas as shown in Fig. 2. Three charging stations have been assigned to four arbitrary nodes in the network since there are no charging stations in place yet. These stations are assumed to be located at junctions 0, 2,3 and 8. Table I lists the network data including the energy consumption ($E$), reliability ($R$), and traveling time in eight-time intervals

{$t_1, t_2, ..., t_8$}. Note that the reliability and traveling time is extracted from [7], and the traveling time is a unit of time.

Applying the results of the proposed algorithms in three different routing options listed in Table II-Table IV. In these tables, the first row is the selected arc, and the second row is the unit of time required for commuting the arc. Travel time in the third row stands for the cumulative unit of time from the origin, and consumption stands for the energy required to traverse the arc. The SOC in the fifth row is the state of the charge, while all the scenarios assumed that the driver starts traveling with the full battery. The last row is the reliability for the arc. In all studies, the driver starts at node 10 and the destination is node 6.

As can be seen in Table II, the shortest path is (10, 9, 8, 7, 6), where the total travel time is 5.75 unit of time, the path reliability is 88.06% and the SOC at the end of the trip is 23%.

The results for the DOT algorithm is demonstrated in Table III, where one can see the lower traveling time in compare with Algorithm 1, with the same reliability and SOC.

TABLE III. RESULTS FOR THE FIFO-DOT ALGORITHM

| Route | 6-7 | 7-8 | 8-9 | 9-10 | Total |
|---|---|---|---|---|---|
| Time | 1.20 | 0.73 | 0.82 | 1.34 | |
| Travel time | 14.8 | 14.07 | 13.25 | 11.92 | 4.8 |
| Consumption | 25% | 17% | 15% | 20% | |
| SOC | 100% | 75% | 58% | 43% | 23% |
| Reliability | 98% | 96% | 96.5% | 97% | 88.06% |

The Non-FIFO behavior of the system is utilized in the last study through proposed Algorithm 3 and the results shown in Table IV. As can be seen, the driver has to stops at nodes 3 and 7, where he had a chance to recharge the battery at node 3. This table has another column for waiting time that indicates the total time that the driver spends at a node with consent. In this case, the total trip time is 4.06 unit of time without considering the waiting time and 8.17 unit of time by considering the waiting time. Additionally, the SOC of 34% and the reliability of 91.29% are comparable with the two other cases mentioned above. Fig. 3 shows the routing result of each algorithm.

TABLE IV. RESULTS FOR THE NON- FIFO WSDOT ALGORITHM

| Route | 6-7 | 7-3 | 3-1 | 1-10 | Total |
|---|---|---|---|---|---|
| Time | 1.20 | 0.69 | 0.69 | 1.48 | |
| Waiting time | 0 | 2.8 | 1.31 | 0 | 4.11 |
| Traveling time | 14.8 | 11.31 | 9.31 | 7.83 | 8.17, 4.06 |
| Consumption | 25% | 15% | 20% | 30% | |
| SOC | 100% | 75% | 74% | 54% | 34% |
| Reliability | 98% | 98% | 98% | 97% | 91.29% |

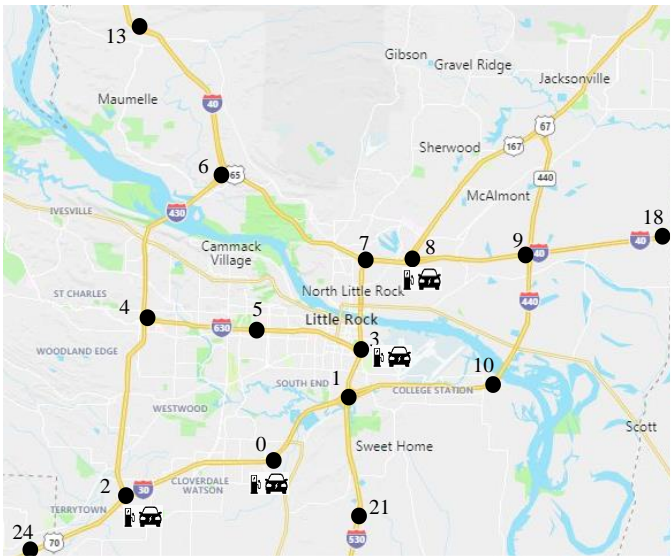

Fig. 2. Central Arkansas highway network.

TABLE I. ENERGY, RELIABILITY, AND TRAVELING TIME FOR THE GIVEN TRANSPORTATION SYSTEM

| ARC | $E$ | $R$ | $t_1$ | $t_2$ | $t_3$ | $t_4$ | $t_5$ | $t_6$ | $t_7$ | $t_8$ |
|---|---|---|---|---|---|---|---|---|---|---|
| 1 | 20 | 97 | 1.48 | 1.88 | 1.48 | 1.34 | 1.34 | 1.34 | 1.34 | 1.08 |
| 2 | 15 | 96.5 | 1.62 | 1.82 | 1.62 | 1.02 | 1.02 | 1.02 | 1.02 | 0.82 |
| 3 | 17 | 96 | 1.13 | 1.13 | 1.13 | 0.73 | 0.73 | 0.73 | 0.73 | 0.73 |
| 4 | 25 | 98 | 1.52 | 1.72 | 1.52 | 1.60 | 1.60 | 1.60 | 1.60 | 1.20 |
| 5 | 30 | 97 | 1.68 | 1.88 | 1.68 | 1.88 | 1.48 | 1.28 | 1.28 | 1.08 |
| 6 | 20 | 98 | 1.14 | 1.54 | 1.14 | 5.76 | 0.69 | 0.74 | 0.54 | 0.34 |
| 7 | 15 | 98 | 1.98 | 2.18 | 1.98 | 0.98 | 0.98 | 0.69 | 0.98 | 0.78 |
| 8 | 20 | 98.5 | 1.97 | 1.97 | 1.97 | 1.57 | 1.17 | 0.97 | 0.97 | 0.97 |
| 9 | 20 | 98 | 3.06 | 3.06 | 3.06 | 1.46 | 1.06 | 0.86 | 0.80 | 0.66 |
| 10 | 25 | 97.5 | 1.87 | 1.87 | 1.87 | 2.07 | 1.67 | 1.67 | 1.47 | 1.47 |
| 11 | 15 | 99 | 1.67 | 1.87 | 1.67 | 1.67 | 1.47 | 0.87 | 0.87 | 0.87 |
| 12 | 20 | 99 | 1.84 | 1.84 | 1.84 | 1.64 | 1.44 | 0.84 | 1.04 | 1.04 |
| 13 | 35 | 99 | 2.52 | 2.52 | 2.52 | 1.72 | 1.72 | 1.52 | 1.40 | 1.32 |

TABLE II. RESULTS FOR THE FIFO ALGORITHM

| Route | 10-9 | 9-8 | 8-7 | 7-6 | Total |
|---|---|---|---|---|---|
| Time | 1.48 | 1.62 | 1.13 | 1.52 | -- |
| Travel time | 0 | 3.1 | 4.23 | 5.75 | 5.75 |
| Consumption | 20% | 15% | 17% | 25% | -- |
| SOC | 100% | 80% | 65% | 48% | 23% |
| Reliability | 97% | 96.5% | 96% | 98% | 88.06% |

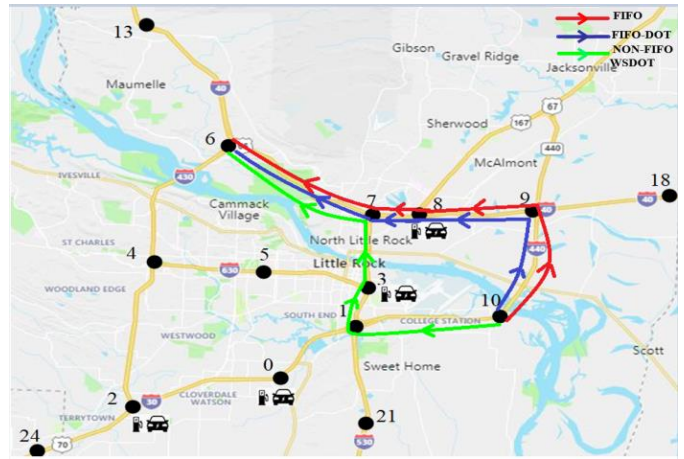

Fig. 3. The routing result of all three routing algorithms

## 6. CONCLUSION

Accounting for en-route delays and alternate routings, it is ATIS networks often exhibit non-FIFO behavior—drivers who depart the origin earlier may not arrive ahead of those who depart later. Having this property in mind, drivers' dynamic routing decisions are modeled in a time-dependent network (with access to waiting en-route data). The traditional time-dependent shortest-path algorithm is extended by capitalizing on EVs feature, which is a requirement for visiting charging stations and relatively long charging times. By taking advantage of this battery feature, a heuristic algorithm is developed to cater for not only the shortest travel time but also the least risk and charging en-route. For further research on this subject, we suggest hiring heuristics and meta-heuristics algorithms [17-19], Genetic Algorithm [20], and machine learning [21,22] to find real-time and near optimum solutions. The exact optimization approach [23-26] can be also pursed to reach an optimum solution. To have more idea about power grid related topics references [27,28] are recommended, and to purse utilizing cognitive approaches to solve the discussed problem in an collaborative environment references [29,30] are recommended.